# Study on dielectric behavior of Lithium Tantalate(LT) nano particle filled poly (vinylidene fluoride) (PVDF) nano composite

S. Satapathy[*a], P. K. Gupta[a], K. B. R. Varma[b], Pragya Tiwari[c] and V. Ganeshan[d]

[a]Laser Materials Development & Device Division, Raja Ramanna Centre for Advanced Technology, Indore 452013, India

[b]Materials Research Centre, Indian Institute of Science, Bangalore 560 012, India

[c]Synchrotron Utilization & Materials Research Division, Raja Ramanna Centre for Advanced Technology, Indore 452 013, India

[d]UGC-DAE Consortium for Scientific research, Khandwa Road, Indore452017, India

[*]Author to whom correspondence should be addressed;

electronic mail: srinu73@cat.ernet.in

Tel.: +91 731 2488655

Fax: +91 731 2488650

**Abstract**

For pyroelectric detector application materials should have low dielectric constant, high pyroelectric coefficient, large non volatile polarization at small applied electric field and low specific heat. Large field (greater than 1200kV/cm) is need to pole ferroelectric polymer poly (vinylidene fluoride) (PVDF) and it has low sensitivity compared to other pyroelectric materials. To increase non volatile polarization at low poling field and to increase pyroelectric coefficient, $LiTaO_3$ (LT) nano particles were added to PVDF matrix to make LT/PVDF composite. It is important to study the dielectric properties of the composite (to be used in detector application) because dielectric constant varies with volume fraction of filler and with frequency. Nano composite films of LT/PVDF with different volume fraction (i. e $f_{LT}$ = 0.047, 0.09 and 0.17) of LT were prepared by dispersing LT nano particles in solution of PVDF. The dielectric properties of LT/PVDF composite were studied by varying the volume fraction of LT. The dielectric permittivity of LT/ PVDF composites increased compared to PVDF as the volume fraction of LT increases but the loss tangent is almost constant at higher frequency. In low frequency region, for $f_{LT}$ = 0.17 the dielectric permittivity of composite is greater than PVDF and LT. The dielectric loss tangent is also increased from 0.04 to 0.175 as $f_{LT}$ increases from 0 to 0.17 at 1 kHz. The dielectric permittivity behavior of composite has been explained using percolation model and space charge polarization model.

# 1. Introduction

For construction of pyroelecric detector, the materials should have large pyroelectric coefficient, high resistivity, surface uniformity, high Curie temperature, a small dielectric constant, and low thermal capacity. The thermal conductance (G) must be as small as possible [1]. The material should produce large non volatile polarization with application of low poling field.

Considerable attention has been devoted in the past decade to study piezoelectric and pyroelectric properties of PVDF because of its favorable properties as a transducer component material. The unique properties of PVDF transducers include low density, low acoustic impedance, wide bandwidth, flexibility, toughness, and ease of fabrication into complex patterns and arrays. These properties are ideally suited to yield a great variety of designs for sensors and actuators in many different fields of application at a potentially low cost [2]. At application of 1200KV/cm the remanent polarization of PVDF is 5μC/cm$^2$ at 10$^{-3}$ Hz and at 20$^0$C [3]. The reported dielectric constant (ε) is 9 at 60Hz and the reported pyroelectric coefficient ($p$) varies from 9.0 * 10$^{-10}$ C cm$^{-2}$/ K [4, 5] to 2.0*10$^{-10}$ C cm$^{-2}$/ K [6, 7]. For achieving high figure of merit ($F = \dfrac{p}{c\varepsilon K}$), the materials having high pyroelectric coefficient ($p$), low dielectric constant (ε) and low K (thermal conductivity) are desirable, since these result in relatively large changes in voltage and temperature respectively [8].

But The PVDF has low value of pyroelectric coefficient ($p$). High field, greater than 1200kV/cm, is needed to pole the PVDF film. That is why these above parameters degrade the performance of PVDF thin film for pyroelectric sensors performance. Due to low value of $p$, its sensitivity is also low. Its poor heat dissipation makes it highly vulnerable to thermal damage.

Lithium tantalate exhibits unique electro-optical, pyroelectric and piezoelectric properties combined with good mechanical and chemical stability and, wide transparency range and high optical threshold. It is rugged, no hygroscopic, and chemically stable. This single crystalline material has become a "workhorse" for modern high-performance pyroelectric applications. In face electrode configurations it has demonstrated both the highest sensitivity and the highest frequency response among commercially practical pyroelectric materials. With a Curie temperature above 600 $^0$C [9], it is resistant to depoling and exhibits linear response throughout a wide operating range [10]. The required coercive field ($E_c$) to pole the LT (congruent) is 200KV/cm and for stoichiometric composition of LT the required field is 17kV/cm [11]. The dielectric constant ($\varepsilon$) of LT nano particle is around 40 [12]. The pyroelectric coefficient of LT is 2.3 * 10$^{-8}$ C cm$^{-2}$/ K [10] which is greater than PVDF.

To reduce poling field of PVDF and to increase pyroelectric coefficient, there is a need to make composite of PVDF with a filler material which having very high value of pyroelectric coefficient, low poling field and dielectric constant comparable to that of PVDF. So the LT nanoparticles are chosen as filler in PVDF matrix, to increase the nonvolatile polarization at small poling filed and pyroelectric coefficient of PVDF.

But in composite the dielectric constant is different from the dielectric constant of the filler and polymer matrix. For sensor application the dielectric constant and filler fraction should be optimized to get low dielectric constant and high non volatile polarization. Composites were prepared using nano fillers of PZT [13], PMN-PT [14],

BaTiO$_3$ [15], carbon nanotubes [16] etc.. But due to addition of PZT and PMN-PT, the non volatile polarization polymer increases as well as the dielectric constant increases due to high dielectric constant of the filler, which is not desireable for detector application. At different situation different models were established to study the dielectric properties of composite. Models like Maxwell-Wagner [17, 18], Logarithimic, Bruggeman [19-25] and Vo-Shi [26-28] are used to describe the dielectric behavior of dielectric mixtures including polymer composites. The enhancement of dielectric constant in percolative stage was explained according to power law [29, 19].

In this report the change in microstructure and dielectric permittivity of LT/PVDF composite with volume fraction of LT have been investigated. The work also focuses on the change in dielectric behavior over broad range of the frequency and volume fraction of LT. The variation of dielectric constant with volume fraction of LT has been explained by dividing total frequency range into low (100Hz to 10$^4$Hz) and high frequency regions (10$^4$Hz to 10$^6$Hz). For $f_{LT}$ = 0.17, the dielectric permittivity of composite is greater than the dielectric constant of pure LT in low frequency range. The dielectric constant in low frequency range was fitted to Bruggeman mixing rule and percolation model. Using the space charge polarization and capacitor model, the variation dielectric constant with volume fraction and frequency of composite was described.

## 2. Experiment

Spherical well separated nano particles (20-40nm) of LT were prepared by adding oleic acid (capping ligand) to alkoxide solution of the LT [12]. The nano particles of LT were dispersed in solution of PVDF and Dimethyl sulphoxide (DMSO). The LT/PVDF composites were cast on glass plates at same temperature. The free standing composite thin films of thickness 30μm were annealed at $90^0$C for 5 hours. The compositions of composite were studied using X-ray diffraction. The microstructure of composite films was examined by scanning electron microscopy (SEM) and atomic force microscopy (AFM). The dielectric measurements were carried out using precision impedance analyzer (HP 4194A) in frequency range of 100Hz to $10^6$Hz.

## 3. Results and discussion

The X-ray diffraction pattern of pure PVDF is shown in Fig. 1(a). The peak at 20.7(2θ value) corresponds to (200) of β-phase PVDF. Fig. 1(b) - (d) show the diffraction pattern of LT/ PVDF composites for $f_{LT}$ = 0.047, 0.09 and 0.17 respectively. Peaks at 23.7, 32.8, 34.7, 39.2, 42.6 and 48.6 correspond to planes (012), (104), (110), (006), (202) and (024) respectively, confirm the proper composition of LT nano particles belong to R3c space group (JCPDS file No. 29-0836).

Fig. 2 shows microscopic pictures of pure PVDF and LT /PVDF composites (for $f_{LT}$ = 0.047, 0.09 and 0.17). In pure PVDF the crystalline regions are separated by non crystalline regions (fig. 2(a)). From AFM it is clear that the crystalline regions of size 400nm are separated from each other by noncrystalline regions. After addition of nano particles to PVDF, there is formation of small regions in which nano particles are embedded by polymer matrix (fig. 2(b)). The size of small regions is around 600nm in case of LT/PVDF composite for ($f_{LT}$ =0.047). Fig. 2(c) and fig. 2(d) show SEM and AFM pictures of LT/ PVDF nano composites for $f_{LT}$ = 0.09 and 0.17 respectively. As shown in the figure the size of regions in which nano particles are embedded are decreases with increase of volume fraction. AFM pictures show that the nano particles come closer and closer as the volume fraction increases. As the volume fraction of LT increases from 0.047 to 0.17, the nanoparticles are coming closer to form a network along all directions.

The dielectric behavior of LT/PVDF nano composites are shown in fig. 3. The dielectric behavior of composite at different volume fraction of LT has been divided in two frequency ranges. Fig. 3 (a) shows the dielectric behavior at low frequency range i. e. from 100Hz to $10^4$Hz and fig. 3 (b) shows the dielectric behavior at high frequency range i. e. $10^4$Hz to $10^6$Hz. Similarly the dielectric loss tangent of composite at different volume fraction of LT within frequency range 100Hz to $10^4$Hz has been shown in fig. 3 (c). The dielectric loss tangent at high frequency range has been shown in fig. 3 (d).

The values of the dielectric permittivity and loss tangent (tan δ) of pure PVDF at $30^0$C are 20.26 and 0.04 respectively at 1 kHz. In case of pure PVDF the dielectric permittivity decreases continuously with increase in frequency at room temperature (fig. 3(a)). But after 10 kHz the dielectric permittivity is almost constant and doesn't vary with frequency. The loss tangent decreases rapidly from 0.05 at 100Hz to 0.04 at 1 kHz, but increases continuously to 0.11 at 1MHz.

For $f_{LT}$ = 0.047, the dielectric permittivity and dielectric loss of composite follow the same behavior as that of PVDF. The values of dielectric permittivity and loss tangent are 24 and 0.07 respectively at 1 kHz. For $f_{LT}$ = 0.047, the loss tangent decreases rapidly from 0.09 at 100Hz to 0.05 at 30kHz, but increases continuously after 30kHz to 0.11 at 1 MHz.

Similarly for $f_{LT}$ = 0.09, the dielectric constant is increased to 28.8 at 1 kHz. The loss tangent is decreased rapidly from 0.09 at 100 Hz to 0.05 at 34 kHz. When $f_{LT}$ is increased to 0.17, then the dielectric constant is increased to 48 at 1 kHz. The loss tangent decreases rapidly from 0.23 at 100 Hz to 0.06 at 66 kHz.

From fig. 3 (b) it is clear that the dielectric constant is independent of frequency at higher frequency range for all volume fraction of composite. But the value of dielectric constant increases with volume fraction of LT. Upto $f_{LT}$ = 0.17, the dielectric constant of composite is in between dielectric constant of polymer and dielectric constant of the LT nano particle within frequency range $10^4$Hz to $10^6$Hz i. e. $\varepsilon_{plymer} < \varepsilon_{composite} < \varepsilon_{filler}$. Within this frequency range the loss tangent of composite is almost same as that of pure PVDF. So from high frequency dielectric constant graph it is clear that the increase in dielectric constant with volume fraction of LT is due to increase in dipolar polarization. Since the dielectric constant of LT nano particle is greater than PVDF, we observed increase in dielectric constant with volume fraction of LT at high frequency range.

But at lower frequency range the dielectric constant of composite strongly depends on frequency. Between 100Hz and 1 kHz the dielectric constant of composite increases sharply with decrease in frequency for all volume fraction of LT compare to other frequencies. At 100Hz, the dielectric constant of LT/PVDF composite increases from 22.1 to 68.4 as volume fraction of LT increases from 0.0 to 0.17. Similarly for 1 kHz the dielectric constant increases from 20.2 to 48 as the volume fraction of LT increases from 0.0 to 0.17. For $f_{LT}$ = 0.17, the dielectric permittivity of LT/PVDF composite is greater than the dielectric constant of the LT nano particles (~40) and dielectric permittivity of polymer (~20) in frequency range 100Hz to 1kHz. From high frequency dielectric analysis it is clear that the increase in dielectric constant is attribute to increase in dipolar contribution which arises from LT nano particles and the dielectric constant of composite is less than dielectric constant of nano particles. But in low frequency range in addition to the dipolar contribution there must be some other polarization contributes to increase in dielectric constant which also depends on the frequency. Since the dielectric constant depends on frequency, the possibility of existence of inter-phase has been ruled out [30]. So the extra polarization arises from space charge polarization.

The PVDF is polarized when an electric field is applied on the polymer. The quantity of accumulate charge depends on the polarity of the polymer. When LT nanoparticles are added to PVDF, then the polarization of composites have been increased compared to PVDF due to the dipolar contribution of LT. But in low frequency region in addition to polarization due to PVDF and LT nano particles, the space charge polarization plays a major role to increase dielectric constant of composite. The space charge polarization arises from the LT/ PVDF interfaces.

Fig. 4(a) and fig. 4(b) show the plot of the dielectric permittivity and the loss tangent of composites against different volume fraction of LT respectively at 1 kHz (low frequency range) and 100 kHz (high frequency range). First consider the plot of dielectric constant and loss tangent with respect to volume fraction at 100 kHz. As the volume fraction of LT is increased in the composite, the dielectric permittivity of composite is also increased. But the loss tangent is almost constant for all volume fraction of LT at 100 kHz i. e. 0.05. So the dielectric constant increase at 100 kHz for all volume fraction of LT compared to PVDF is attributed to the increase in dipolar contribution from LT.

But the dielectric constant and loss tangent both increases with volume fraction of LT at 1 kHz. As per discussion for fig.3, the space charge polarization due to LT/PVDF interface contributes to dielectric constant in addition to polarization contribution from LT and PVDF. The increase in dielectric loss with volume fraction of LT at 1 kHz supports the fact of the space charge polarization contribution. As the volume fraction increases from 0.0 to 0.17 the dielectric constant of LT/PVDF composite increases from 20.2 to 48. For $f_{LT}$ = 0.17, the dielectric permittivity of LT/PVDF composite is greater than the dielectric constant of the LT nano particles (~40) and dielectric permittivity of polymer (~20). To solve the dielectric properties of composite at low frequency we took help of percolation model.

According to Maxwell mixing rule for composite, the matrix is considered as an uniformly polarized medium surrounding the sphere (filler) under consideration. For dilute spheres Maxwell calculated $\varepsilon_c$ (dielectric constant of composite) by taking the result of one sphere and multiplying by the sphere number density [31].

The Bruggeman improved the result of Maxwell by essentially iterating the polarization of the sphere and the matrix until the average net polarization vanishes- it is a self consistent field calculation [31].

This results in following equation.

$$\left(\frac{\varepsilon_1 - \varepsilon_c}{\varepsilon_1 + 2\varepsilon_c}\right)\phi + \left(\frac{\varepsilon_2 - \varepsilon_c}{\varepsilon_2 + 2\varepsilon_c}\right)(1-\phi) = 0 \qquad (1)$$

where $\varepsilon_1$, $\varepsilon_2$ and $\varepsilon_c$ are dielectric constant of filler, matrix and composite respectively. The "$\phi$" is volume fraction of the filler. This expression treats the phases in symmetric fashion. Without considering space charge polarization the Bruggeman model was fitted to our experimental result at 1 kHz. There is large discrepancy between experimental results and the results obtained from model (curve2, fig. 4(c)). The dielectric constant of a material is a function of capacitance, which is also proportional to the quantity of charge stored on either surface of the sample under an applied field. The dielectric constant of composite increases with addition of LT reflects the formation of capacitance network of LT nano particles. As the volume fraction of LT increases capacitance network also increases. The increase in dielectric constant of composite above the dielectric constant of LT and increase in loss with volume fraction of LT indicates the conducting network of tiny capacitors. So the increase in dielectric constant with volume fraction of LT can be described using percolation model. In percolative stage, the enhancement of dielectric permittivity can be explained according to the following power law [19, 29]:

$$\varepsilon = \varepsilon_2 \left( \frac{f_c - f_{fill}}{f_c} \right)^{-q} \qquad (2)$$

where $\varepsilon_2$ is the dielectric constant of PVDF matrix, $f_{fill}$ is the volume fraction of the filled particles, $f_c$ is the percolation threshold and q is a critical exponent of about 1.

The experimental dielectric permittivity data have been fitted with the data generated using percolation model. The experimental dielectric permittivity values are in good agreement with the theoretical values obtained using power law (eq.$^n$ 2) (curve1, fig. 4(c)). The dielectric constant has been plotted w. r. t. $(f_c - f_{LT})$ in fig. 4(d), which is a straight line. The values of $f_c$ and q obtained from plot are 0.29 and 0.98 respectively. Since filler particles in LT/PVDF composite are oxide particles, the percolation threshold ($f_c \approx 0.29$) of LT/PVDF composite is greater compared to percolation threshold ($f_c \approx 0.16$) of composite in which fillers are conducting spherical particles. So for $f_{LT} = 0.17$, the percolation in composite is about to start but well below the percolation threshold. The increase in dielectric permittivity can be understood by gradually assessing the formation of micro capacitance networks in the LT/PVDF composites when the concentration of LT increases. The increase in dielectric loss tangent in low frequency region can be described by formation of conducting path through capacitance network.

## 4. Conclusion

The dielectric permittivity of the LT/PVDF composite is increased with increase of volume fraction of LT in PVDF matrix. In high frequency range the dielectric loss is almost constant for all volume fraction of LT in LT/PVDF composite but the dielectric constant increases with volume fraction of LT. The dielectric constant doesn't depend on frequency. So the increase in dielectric constant is due to the increase in dipole moment due to addition of LT particles. In low frequency region the dielectric constant depends on frequency and loss also increases as volume fraction of LT increases. The dielectric constant of composite also increases beyond the dielectric constant of LT filler, which is greater than PVDF, in low frequency range. So the space charge polarization in addition to dipolar polarization due to PVDF and LT contributes to the dielectric constant in low frequency range. The space charge arises at the LT/PVDF interface. As the LT fraction increases the interfacial charge also increases. SEM shows uniform distribution of nano particles in polymer matrix. For $f_{LT} = 0.17$, the dielectric permittivity of composite is increased to 48 at 1 kHz which is greater than the dielectric permittivity of PVDF and LT. The increase in dielectric constant attributes to the increase in the quantity of the accumulated charge at LT/PVDF interfaces in frequency region. The increase in dielectric permittivity with higher volume fraction of LT has been described using percolation theory in low frequency range. So it is found that the dielectric constant of LT/PVDF composite is less than the dielectric constant of LT nano particles for all volume fraction of LT (upto 0.17) in high frequency region. Low frequency region dielectric constant increases beyond the dielectric constant of nano particles due to space charge polarization and capacitance network formation. But the dielectric constant increases only upto 48 at 1 kHz for $f_{LT} = 0.17$ which is very less compared to PZT/PVDF and PMN-PT/PVDF nano composites. According to percolation model the percolation threshold for LT nano particle is found to be 0.29. So after volume fraction 0.29, the sample will be conducting. In our case the volume fraction of LT is 0.17 which is well below the percolation threshold value. So for dielectric point of view the LT/PVDF can be used in pyroelectric sensor.

**Caption of figures**

**Fig. 1.** X-ray diffraction patterns of (a) pure PVDF and of LT /PVDF composites for $f_{LT}$ = (b) 0.047 (c) 0.09 and (d) 0.17.

**Fig. 2.** SEM and AFM micrographs of (a) pure PVDF and of LT/PVDF composites for $f_{LT}$ = (b) 0.047 (c) 0.09 and (d) 0.17.

**Fig. 3.** Dependence of dielectric permittivity of composite on frequency at room temperature (a) for low frequency range (100Hz to $10^4$Hz) (b) for high frequency range ($10^4$ Hz to $10^6$Hz). Dependence of dielectric loss tangent of composite on frequency at room temperature (c) for low frequency range (100Hz to $10^4$Hz) (d) for high frequency range ($10^4$ Hz to $10^6$Hz).

**Fig. 4.** Variation of (a) dielectric permittivity and (b) dielectric loss tangent of LT/PVDF composites with volume fraction of LT at 1kHz and 100kHz. (c) The dielectric permittivity of LT/PVDF composites at 1 kHz is fitted to Bruggeman (curve 2), and percolation model (curve 1). (d) The dielectric permittivity of LT/PVDF composite has been plotted against $(f_c - f_{LT})$.

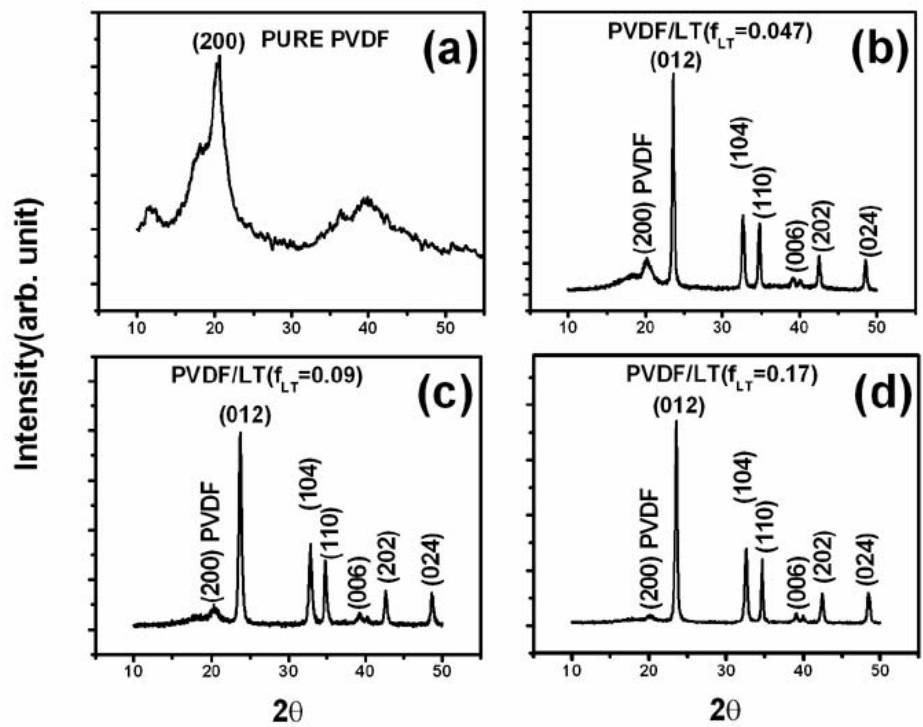

**Fig.1.**

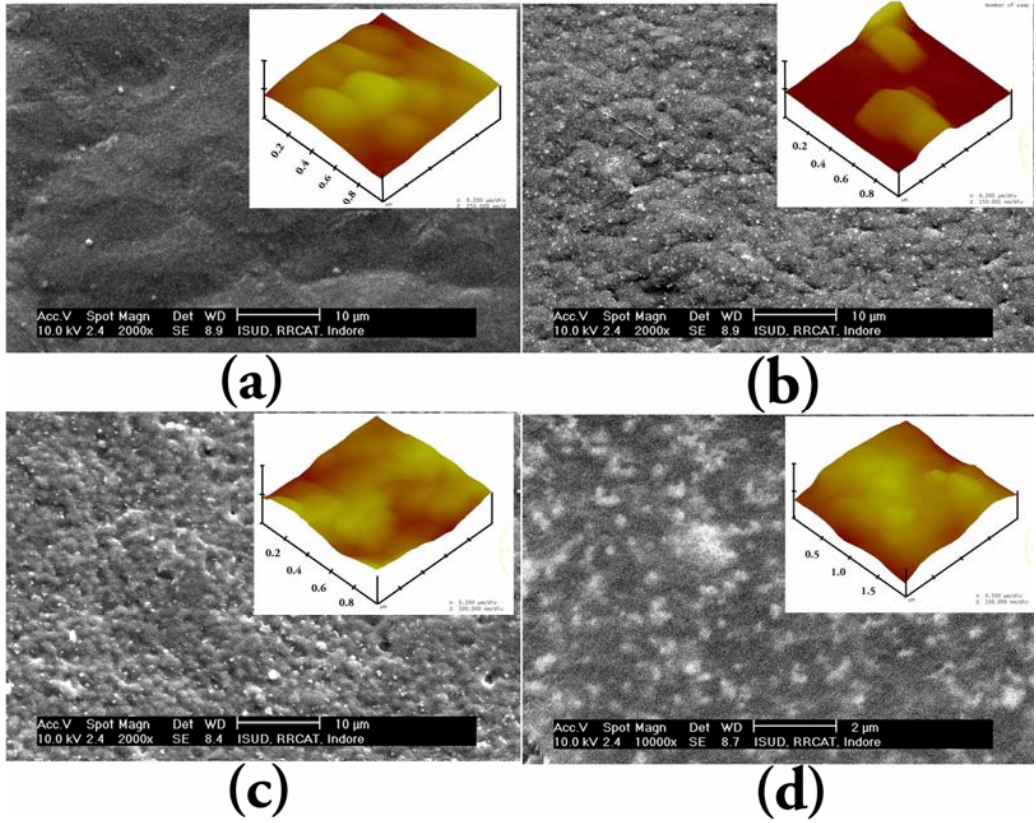

**Fig.2**

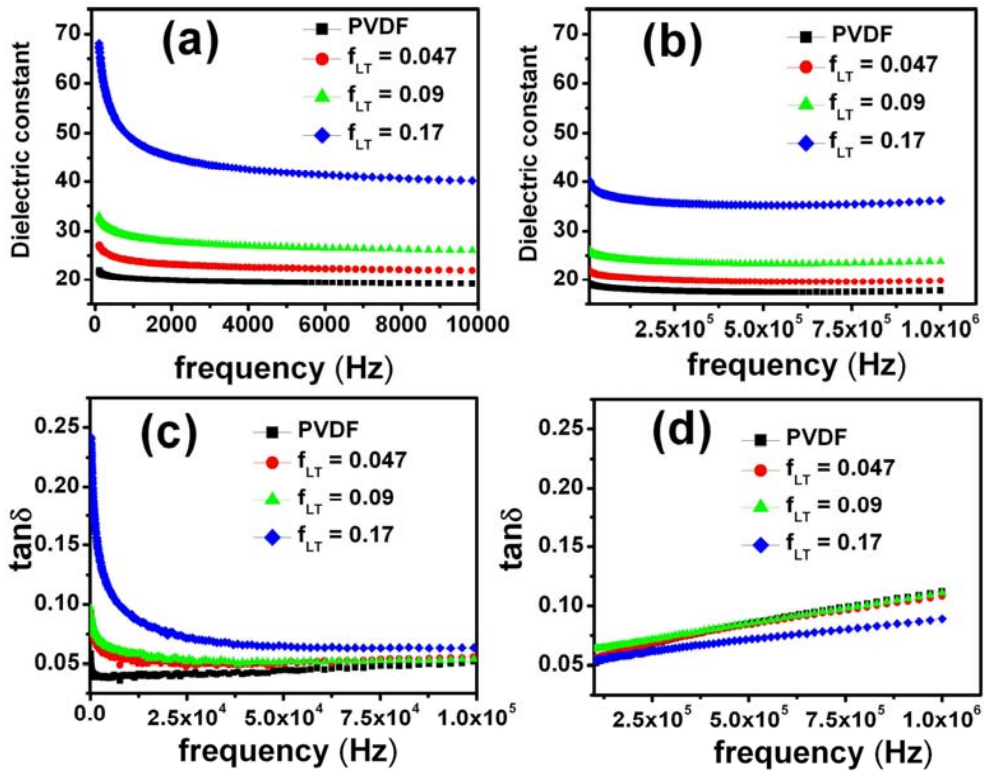

**Fig.3**

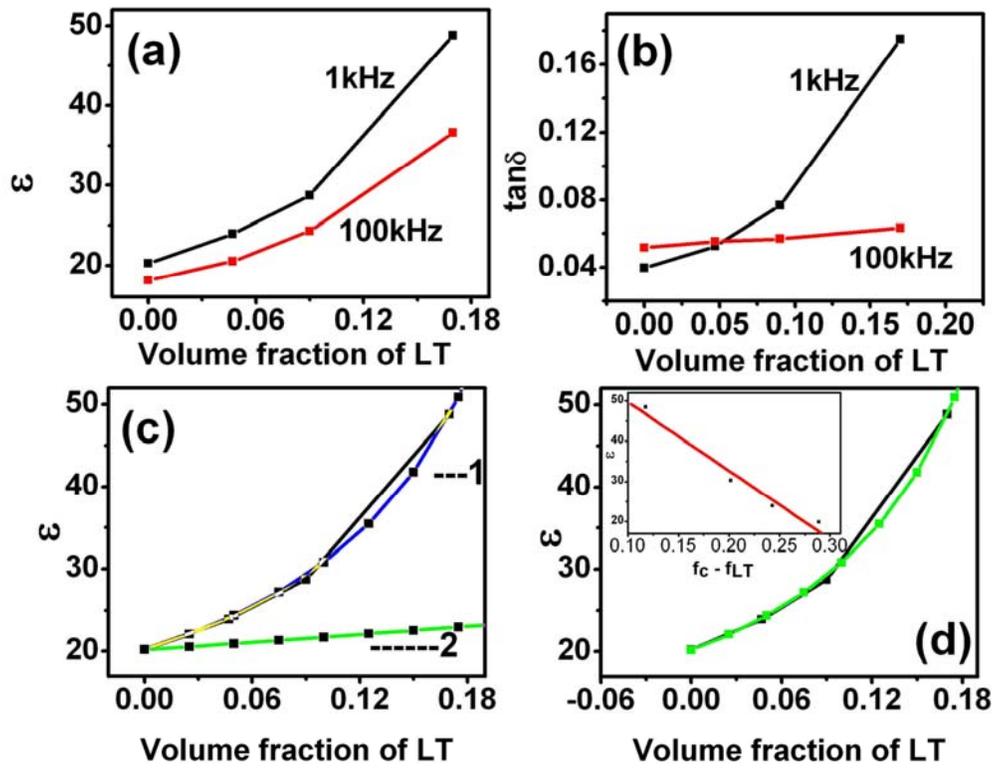

**Fig.4**